\documentclass[journal,twoside,web]{ieeecolor}
\usepackage{booktabs}
\usepackage{multirow}
\usepackage{generic}
\usepackage{cite}
\usepackage{amsmath,amssymb,amsfonts}
\usepackage{algorithmic}
\usepackage{graphicx}
\usepackage{algorithm,algorithmic}
\usepackage{hyperref}
\hypersetup{hidelinks}
\usepackage{textcomp}
\def\BibTeX{{\rm B\kern-.05em{\sc i\kern-.025em b}\kern-.08em
    T\kern-.1667em\lower.7ex\hbox{E}\kern-.125emX}}
\begin{document}
\bstctlcite{IEEEexample:BSTcontrol}

\title{Towards Scalable Image Guidance: Multimodal AI for Real-Time Motion Estimation Combining X-Ray and Physiological Signals}

\author{
Chuhan Wang, Zhuang Xiong, Mark Gardner, Gregory Willson,
Brett Clark, Frances Kan, Vicky Chin, Trevor Moodie,
Tim Wang, Jeffrey Barber, Tony Barry,
Eric Hau, Xingzhou Liu, Poornima Balaji, Pierre Qian, Vu Toan Tran, Juntang Lu, Saurabh Kumar, Jeremy Booth, Ricky O'Brien,
Paul Keall, and Chandrima Sengupta
\thanks{The authors sincerely thank all patients who participated in this study. The authors also thank James Grover, Michael Ferraro and Helen Ball from Image X for their internal review and constructive feedback and Julia Johnson for assistance with Fig. 1. The authors would like to thank A/Prof Verity Ahern from Westmead Radiation Oncology for her oversight of the cardiac animal study and LARK clinical trial, and Dr Yoo Young Lee, Tiarne Murray, and Andrew Hickey for their assistance with data acquisition in both studies. The authors also acknowledge the Radiation Oncology Departments at the Crown Princess Mary Cancer Center and Royal North Shore Hospital for their clinical support of this study.
Chandrima Sengupta is supported by the Cancer Institute NSW Early Career Fellowship (2024/ECF1592). Paul Keall is supported by the Australian Government National Health and Medical Research Council (2041792). The research program was supported by the NHMRC Ideas grant 2046686, Medical Research Future Fund grant NCRI000139, NHMRC Synergy grant 2018592.}
\thanks{Chuhan Wang, Zhuang Xiong, Mark Gardner, Gregory Willson, Brett Clark, Frances Kan, Vicky Chin, Paul Keall, and Chandrima Sengupta are with the Image X Institute, School of Health Sciences, Faculty of Medicine and Health, The University of Sydney, Sydney, NSW, Australia (e-mail: chuhan.wang@sydney.edu.au; chandrima.sengupta@sydney.edu.au).}
\thanks{Brett Clark is also with the Department of Radiation Oncology, Peter MacCallum Cancer Center, Melbourne, VIC, Australia.}
\thanks{Trevor Moodie, Tim Wang, Jeffrey Barber, and Eric Hau are with the Radiation Oncology Network, Westmead Hospital, Sydney, NSW, Australia.}
\thanks{Tony Barry, Xingzhou Liu, Poornima Balaji, Pierre Qian, Vu Toan Tran, Juntang Lu, and Saurabh Kumar are with the Cardiology Department, Westmead Hospital, Sydney, NSW, Australia.}
\thanks{Jeremy Booth is with Radiation Oncology, Royal North Shore Hospital, Sydney, NSW, Australia.}
\thanks{Ricky O'Brien is with the School of Health and Biomedical Sciences, RMIT University, Melbourne, VIC, Australia.}
\thanks{Corresponding author: Chandrima Sengupta (e-mail: chandrima.sengupta@sydney.edu.au).}}

\maketitle

\begin{abstract}
Real-time image-guided radiotherapy using specialized MRI-Linac systems, which enable visualization of tumor motion and adaptation of treatment delivery, has demonstrated approximately 50\% reductions in treatment-related toxicity. However, high costs limit adoption, despite 70\% of radiotherapy centers wanting real-time image-guidance. We propose a scalable and resource-efficient solution integrating multimodal signals from standard X-ray-based radiotherapy and interventional imaging equipment, enabling real-time image guidance without specialized hardware. We developed the \underline{S}patial and \underline{P}hysiological \underline{A}daptive \underline{R}elevance \underline{Net}work (SPAR-Net), which combines a pre-trained RAD-DINO encoder for extracting anatomical features from X-ray images with a Mamba-based encoder for modeling temporal dependencies in physiological signals, including ECG and respiratory signals. To enable motion-aware fusion, we introduce a Masked Physiology-guided Spatial Relevance Unit (MPSRU) and a Channel Relevance Unit (CRU). The MPSRU combines training-only random image feature masking with cross-attention to identify regions associated with physiological motion, while the CRU regulates physiological contributions across feature channels. Together, these units enable selective physiological guidance while preserving image-derived anatomical information under imaging distribution shifts. The framework is evaluated on cardiac, lung and liver tumor segmentation tasks involving complex motion patterns and imaging conditions. SPAR-Net demonstrated competitive performance across segmentation metrics, with the greatest improvements observed in HD95. Mean HD95 values were 8.7, 2.7 and 3.2 mm for the three anatomical sites, representing reductions of 40\%, 10\% and 6\%, respectively, compared with the strongest competing methods. These results demonstrate that by leveraging routinely available imaging and physiological data, SPAR-Net offers a promising approach toward accessible, resource-efficient real-time image guidance. 
\end{abstract}


\section{Introduction}
\label{sec:introduction}
\IEEEPARstart{A}{ccurate} real-time motion estimation of tumors and surrounding critical anatomical structures remains a major challenge in cancer radiotherapy~\cite{ref1,ref2}, cardiac radiotherapy~\cite{ref3} image-guided interventions~\cite{ref4}, where precise targeting is essential for treatment efficacy. This challenge is particularly pronounced in the thorax~\cite{ref5} and abdomen~\cite{ref6,ref7}, where respiratory motion can reach up to 4 cm~\cite{ref6}, while cardiac structures undergo additional intrinsic contractions~\cite{ref8}, resulting in complex motion between targets and surrounding anatomy. Uncorrected motion can compromise treatment accuracy across these applications~\cite{ref4,ref9}, leading to target underdosage and increased irradiation of healthy tissues during radiotherapy, as well as inaccurate instrument navigation during interventional procedures. 

Real-time image guidance~\cite{ref10}, which enables continuous motion monitoring and adaptation during treatment, has halved treatment-related toxicity in prostate cancer radiotherapy~\cite{ref11}, reduced treatment volumes by 30\% in the lung~\cite{ref12} and demonstrated 30\% reduction in healthy liver dose~\cite{ref13}. However, the existing solutions for improved clinical outcomes have been either generated using specialized imaging and treatment systems including MRI-Linacs~\cite{ref11}, limiting widespread accessibility, or require invasive surgical marker implantation~\cite{ref6}. Findings from the European Society of Radiotherapy and Oncology Patterns of Practice for Adaptive and Real-Time Radiation Therapy survey from 200 hospitals from 71 countries highlight this substantial gap between clinical demand for real-time image guidance and the availability of accurate, affordable and scalable technologies~\cite{ref14}. These limitations underscore the need for widely accessible, real-time image-guidance technologies that can be integrated into conventional radiotherapy and interventional imaging systems.

Standard radiotherapy and interventional imaging systems are equipped with gantry-mounted X-ray imagers, providing an accessible means of visualizing internal anatomy during treatment. 

Accurate motion estimation requires effective feature extraction from X-ray images. Deep learning methods based on convolutional~\cite{ref36,ref37}, attention mechanisms~\cite{ref38} and state-space architectures~\cite{ref39}, alongside foundation models~\cite{ref21,ref22,ref23,ref24,ref25} have been increasingly applied to medical image analysis, including task-specific segmentation including 2D and 3D volumetric image segmentation.

However, X-ray-based real-time image-guidance from standard equipment in radiotherapy and interventional radiology presents several challenges: (1) Conventional two-dimensional X-ray imaging is limited by overlapping anatomy, poor soft-tissue contrast, restricted field-of-view and additional radiation exposure. (2) Continuous gantry rotation changes X-ray projection angles, producing substantially different views of the same moving anatomy and complicating robust anatomical tracking. (3) A single X-ray projection may provide insufficient information to distinguish simultaneous respiratory and cardiac motion, particularly when multiple anatomical structures move independently. (4) Accurate anatomical localization must be achieved within 500 ms to enable timely responses to target motion~\cite{ref50}.

Deep learning methods have been investigated for target motion estimation from 2D X-ray images across various anatomical sites, including the prostate~\cite{ref15}, lung~\cite{ref16,ref17}, pancreas~\cite{ref18}, liver~\cite{ref19}, and head and neck~\cite{ref20} utilizing domain adaptation~\cite{ref26} and self-supervised learning~\cite{ref24}. Although these methods can improve the transferability of image-based representations across imaging conditions, these methods often exhibit limited localization accuracy, struggle with restricted anatomical visibility, and have often been validated only on simulated imaging datasets.

To address these limitations, physiological signals including electrocardiography (ECG), respiratory surface motion, electromyography (EMG) and electroencephalography (EEG) may provide complementary information for estimating current and predicting future anatomical motion~\cite{ref55} without additional radiation exposure. Respiratory and ECG signals have been successfully integrated with three-dimensional volumetric imaging including 4D computed tomography (CT) reconstruction and motion modeling~\cite{ref32,ref33} demonstrating their usability for motion estimation. However, integrating physiological signals with 2D X-ray imaging remains challenging for real-time image guidance due to the limited field of view (FOV) and stringent real-time performance requirements.

\begin{figure*}[!t]
    \centering
    \includegraphics[width=0.65\textwidth]{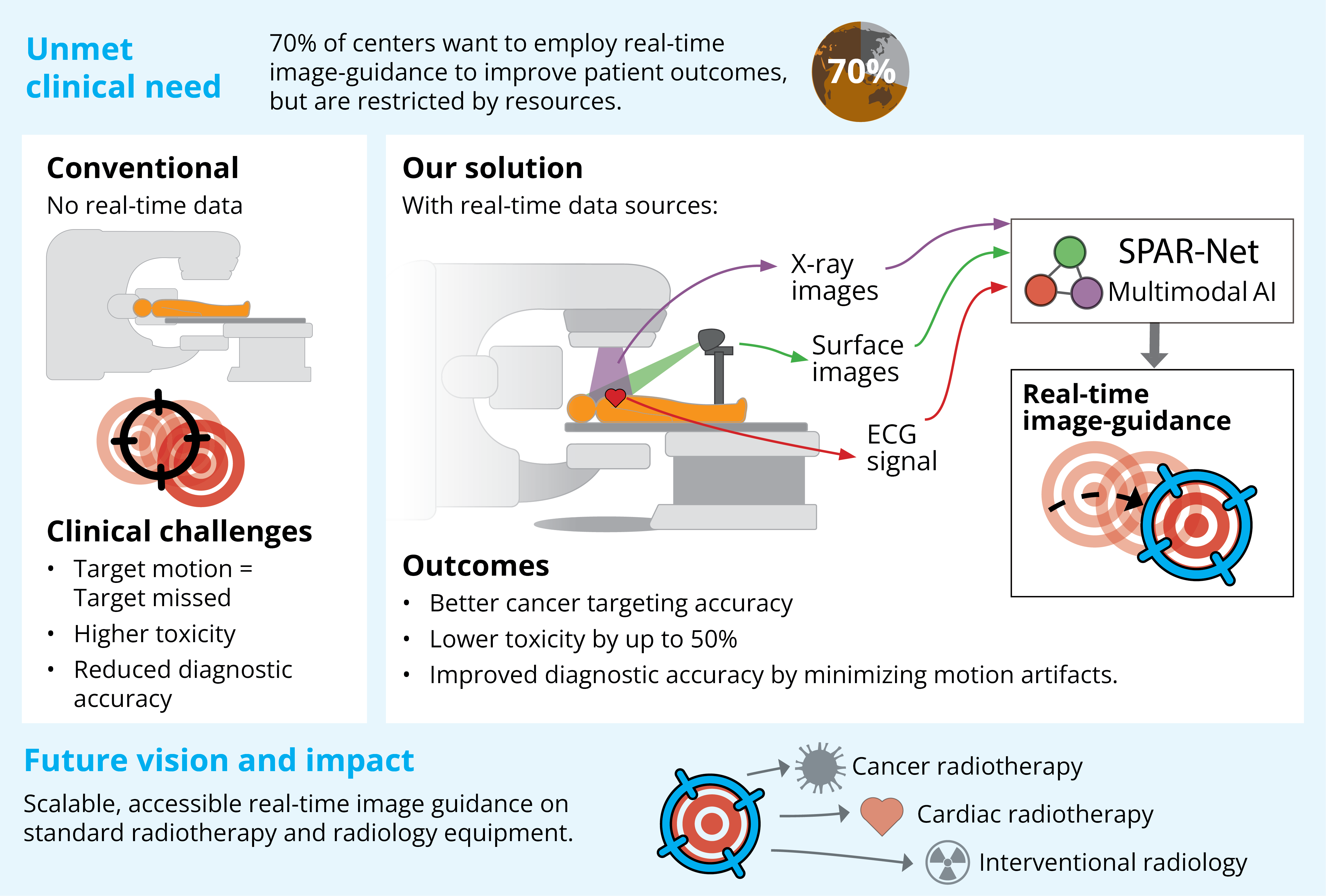}
    \caption{\textbf{Study vision:} One shared challenge, multiple disciplines and a common goal of accessible and scalable real-time image guidance through multimodal AI.}
    \label{fig:framework}
\end{figure*}

ECG and respiratory signals are sequential in nature, and their relationship with anatomical motion depends on temporal context rather than isolated measurements. Recurrent neural networks (RNNs)~\cite{ref27,ref28} and attention-based sequence models~\cite{ref29,ref30} have been widely used to model temporal dependencies in physiological data. More recently, state-space models such as Mamba~\cite{ref31} have provided an efficient mechanism for modeling long-range dependencies in sequential data. However, integrating physiological signals with X-ray imaging presents several challenges, including extracting robust anatomical features from low-contrast X-ray projections under varying imaging conditions, modeling distinct temporal dynamics of ECG and respiratory signals, and selectively incorporating motion-relevant physiological information while preserving image-derived anatomical representations.

\begin{figure*}[!t]
    \centering
    \includegraphics[width=0.90\textwidth]{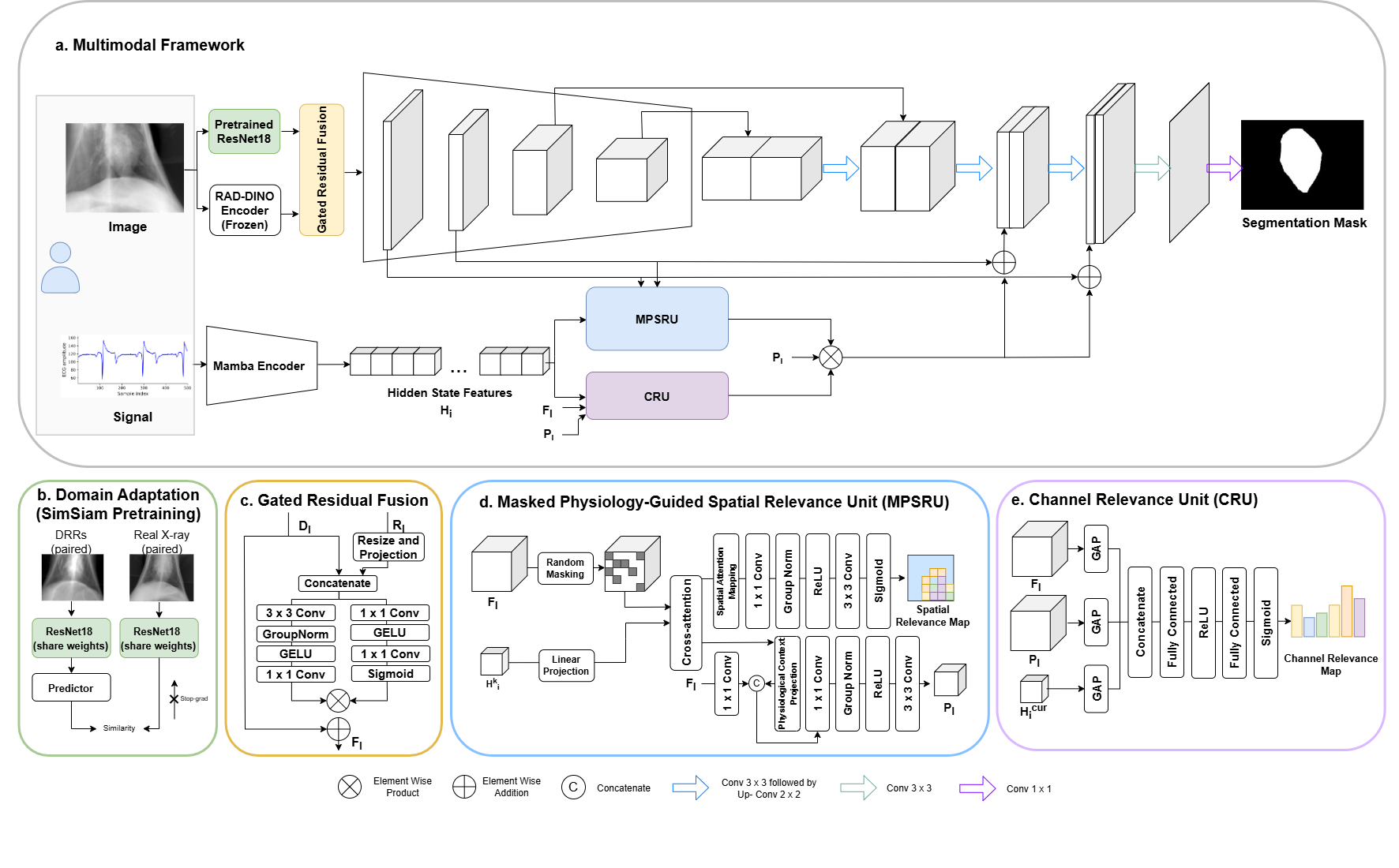}
    \caption{Overview of SPAR-Net. (a) Overall architecture integrating image and physiological signal representations through a U-shaped segmentation network. Image features are extracted using a frozen RAD-DINO encoder, with optional gated residual fusion incorporating features from a SimSiam-pretrained ResNet-18 domain encoder for the cardiac dataset. Physiological signals are encoded using Mamba, and the resulting hidden states guide image features through the Masked Physiology-Guided Spatial Relevance Unit (MPSRU) and Channel Relevance Unit (CRU). (b) SimSiam pretraining of the ResNet-18 domain encoder. (c) Gated residual fusion integrating domain-adaptive and RAD-DINO features. (d) MPSRU estimating spatial relevance through cross-attention between physiological and image features, with random image feature masking applied during training only. (e) CRU estimating channel-wise relevance to selectively modulate physiological guidance across image feature channels.}
    \label{fig:jbhi_fig2}
\end{figure*}

In this study (Fig.~\ref{fig:framework}), we propose \underline{\textbf{S}}patial and \underline{\textbf{P}}hysiological \underline{\textbf{A}}daptive \underline{\textbf{R}}elevance \underline{\textbf{Net}}work (SPAR-Net), a multimodal framework for real-time motion estimation that integrates X-ray imaging with synchronized physiological signals including ECG and respiratory signals Fig.~\ref{fig:jbhi_fig2}. A pretrained encoder, RAD-DINO~\cite{ref24}, is adapted to extract patient-specific imaging characteristics, with an optional ResNet-18~\cite{ref34} domain encoder, pretrained using SimSiam~\cite{ref35}, incorporated when substantial imaging domain shifts are present. A Mamba-based encoder captures the temporal dynamics of ECG and respiratory signals. To selectively incorporate motion-relevant physiological information, we introduce spatial and channel relevance mechanisms that adaptively guide anatomical feature representations. SPAR-Net aims to improve segmentation under varying physiological and imaging conditions. To the best of our knowledge, this is the first study to investigate a unified multimodal AI framework for real-time anatomical motion estimation from three anatomic sites.

The main contributions of our work are summarized as follows: 
\begin{itemize} 
 \item We develop SPAR-Net, a scalable multimodal AI framework that integrates X-ray imaging with synchronized physiological signals to improve the accuracy of motion estimation through target segmentation mask prediction, leveraging complementary anatomical and physiological information using data available from standard imaging equipment.  
 \item We propose two novel modules, the \textbf{M}asked \textbf{P}hysiology-guided \textbf{S}patial \textbf{R}elevance \textbf{U}nit (MPSRU) and the \textbf{C}hannel \textbf{R}elevance \textbf{U}nit (CRU) to adaptively identify where and to what extent physiological information should contribute to anatomical feature representations. The MPSRU combines training-only random feature masking with cross-modal attention to estimate spatial relevance and generate physiology-informed image features. By partially masking image features during training, the MPSRU encourages the model to identify motion-relevant regions using physiological context, reducing dependence on appearance variations associated with changing X-ray projection angles. The CRU further estimates channel-wise relevance by integrating image features and physiological context, allowing physiological guidance to be selectively weighted across feature channels. Together, these modules provide spatially and channel-wise adaptive physiological guidance while preserving image-derived anatomical information.
 \item We evaluate SPAR-Net using a unique dataset of synchronized X-ray imaging and physiological signals across three anatomical sites: cardiac, lung and liver, encompassing diverse motion patterns, imaging conditions and data sources.
\end{itemize}

\section{Related Work}
\label{sec:related work}
\subsection{Deep Learning for X-ray Based Radiotherapy Image Segmentation}
Deep learning has substantially advanced medical image segmentation through the development of increasingly sophisticated feature extraction architectures. Convolutional architectures, including UNet~\cite{ref36} and DeepLabV3+~\cite{ref37}, have demonstrated the effectiveness of encoder–decoder structures and multi-scale feature extraction for medical image segmentation. TransUNet~\cite{ref38} extends these approaches by integrating Transformer-based self-attention to capture global contextual information alongside local features. More recently, U-Mamba~\cite{ref39} has incorporated state-space modeling into convolutional architectures to efficiently capture long-range spatial dependencies in biomedical images.  These methods have been widely applied to 2D and 3D medical images with relatively comprehensive anatomical context. In contrast, radiotherapy X-ray imaging presents additional challenges: the limited FOV restricts anatomical information, while the need to minimize imaging radiation dose constrains image quality, making accurate target segmentation more challenging.

Building upon these general segmentation architectures, patient-specific approaches~\cite{ref15} have integrated generative models, such as conditional generative adversarial networks, with U-Net-based segmentation networks to facilitate markerless tumor localization from treatment-time kilovoltage (kV) X-ray projections. In parallel, a cross-patient markerless localization framework~\cite{ref40} has employed DUCK-Net~\cite{ref41} for lung tumor segmentation, demonstrating the feasibility of X-ray-based localization without patient-specific retraining. Despite these advances, reliable target localization remains challenging due to low soft-tissue contrast, overlapping anatomical structures, restricted fields of view, and variations in projection geometry. These challenges are further compounded by inter-fraction anatomical variations, which can alter target appearance and surrounding structures even within the same patient across different treatment sessions.

\subsection{Foundation Models and Pretrained Encoders in Medical Imaging}
Foundation models have shown considerable potential for medical image segmentation. Medical Segment Anything Model~\cite{ref21} (MedSAM) extends the Segment Anything Model (SAM)~\cite{ref42} to medical imaging, enabling prompt-based segmentation across diverse imaging modalities through large-scale training. However, its training across heterogeneous modalities is not specifically optimised for X-ray imaging. Meanwhile, the success of DINOv2~\cite{ref43} in learning transferable visual representations through self-supervised learning has motivated the development of domain-specific pretrained encoders. RAD-DINO applies image-only self-supervised learning to chest X-rays, demonstrating its effectiveness in downstream classification and segmentation tasks. Similarly, EVA-X~\cite{ref44} combines contrastive learning and masked image modeling to enhance X-ray feature learning, while RayDINO~\cite{ref45} further explores large-scale pretraining and task-specific adaptation across diverse imaging tasks.

Despite their ability to learn rich radiographic representations, pretrained foundation models may exhibit limited performance when directly applied to treatment-time kV X-ray segmentation due to differences in imaging characteristics and anatomical appearance. Furthermore, existing medical imaging foundation models are predominantly pretrained using image-only or image–text data, without explicitly incorporating physiological signals to model their temporal associations with anatomical motion.

\subsection{Temporal Feature Extraction}
Deep learning has been widely investigated for modeling temporal dependencies in physiological signals, including ECG and respiratory measurements. Convolutional Neural Network (CNN) have been employed to extract local temporal patterns, while RNNs and long short-term memory (LSTM) networks~\cite{ref28} capture sequential dynamics through recurrent hidden states. Transformer-based approaches~\cite{ref30} further extend temporal modeling by employing self-attention to capture dependencies across extended sequences. More recently, Mamba introduces selective state-space modeling, enabling efficient representation of long-range temporal dependencies through input-dependent state transitions.

Although these methods can capture temporal dependencies, physiological signals alone do not directly determine anatomical motion. ECG reflects cardiac electrical activity, while external respiratory signals provide indirect estimates of internal motion, with correlations that may vary across patients and treatment sessions~\cite{ref46}. Therefore, combining physiological information with X-ray imaging may improve anatomical localization and motion estimation.

\section{Method}
\label{sec:method}
\subsection{Datasets}
To evaluate the performance of SPAR-Net across thoracic and abdominal anatomical sites with varying motion characteristics and imaging conditions, we used datasets acquired from approved clinical trials and preclinical studies, representing three distinct challenges:  (1) Cardiac, where the position of the whole heart was estimated, accounting for complex motion arising from cardiac-respiratory motion; (2) lung, where respiratory motion of relatively visible tumors was estimated from X-ray projections with the challenge of accurately localizing small targets occupying only a limited region of the image; and (3) liver, where respiratory motion of tumors not directly visible in X-ray projections was estimated. We have used the cardiac dataset as the primary dataset for ablation studies.

\textbf {Cardiac dataset:} Preclinical data from eight sheep were obtained from a sheep study approved by the Westmead Local Health District (approval \#5184), in which sheep underwent cardiac irradiation to investigate the effects of radiation on cardiac arrhythmias in a single treatment session delivering 25 Gy of radiation dose. The dataset included planning four-dimensional CT (4DCT) phase images, contoured structures, kilovoltage (kV) X-ray images and electrocardiography (ECG) signals to investigate complex cardiorespiratory motion. A custom ECG acquisition system was developed because commercially available devices did not provide access to the raw ECG signals required for real-time integration. The system incorporated a Vernier BTA-EKG module and an Arduino Uno microcontroller to record ECG signals at 200 Hz and perform real-time QRS detection using the Pan–Tompkins algorithm. Electrical isolation was implemented to ensure patient safety.

Since the sheep were treated in a single treatment session, for training, digitally reconstructed radiographs (DRRs) were generated~\cite{ref47,ref48,ref49} from the planning CT representing simulated kilovoltage X-ray images evenly spaced every 1 deg. Approximately 50,000 DRRs, including augmented images, were generated per sheep at a resolution of 1024 × 768 pixels. Of the eight sheep, three had time-synchronized ECG and X-ray data available for evaluating SPAR-Net, while the remaining five sheep were used for domain encoder pretraining.

\textbf {Ground truth:} Model evaluation was performed using 500 real X-ray projections acquired during the treatment session per sheep, resulting in 1500 images in the dataset. For ground truth (GT), cardiac contours were manually aligned to each X-ray image by two clinicians to establish reference anatomical positions.

\textbf {Lung dataset:} Lung cancer patient data were obtained from the Markerless Image Guidance Using Intrafraction Kilovoltage X-ray Imaging for Lung Cancer Radiotherapy (VALKIM, NCT04310891) clinical trial. Two patients treated over 4-8 treatment sessions were included. Approximately 500 X-ray projections acquired during the first treatment session were augmented using random flipping, affine transformations, brightness and contrast adjustments, and Gaussian noise, and used for model training, while images from 2-5 treatment sessions on different days were reserved for independent testing, resulting in 7328 real X-ray images in the testing dataset.

\textbf {Liver dataset:} Liver cancer patient data were obtained from the Liver Ablative Radiotherapy using Kilovoltage Intrafraction Monitoring (LARK, NCT02984566) clinical trial. Five patients with liver cancer, treated over 3-5 treatment sessions under free-breathing conditions, were included in this study. Approximately 500 X-ray projections acquired during the first treatment session were augmented using random flipping, affine transformations, brightness and contrast adjustments, and Gaussian noise, and used for model training, while images from 2-4 treatment sessions on different days were reserved for independent testing, resulting in 5538 real X-ray images in the testing dataset.

For both lung and liver datasets, the available data included structure sets, respiratory surface motion signals recorded during treatment delivery, and intratreatment 2D X-ray projection images from all treatment sessions. Respiratory surface signals were acquired from lung and liver cancer patients using the Real-time Position Management (RPM) system, while ECG signals were not collected for these patients.

\textbf {Ground truth:} For lung and liver patients, implanted gold fiducial markers provided reference tumor positions for evaluating model performance. The fiducial markers were masked in the X-ray images to prevent the model from using them for motion estimation. This ensured that the model relied on anatomical features rather than implanted markers, while the marker positions could still be used as GT for evaluating motion estimation accuracy.

\subsection{Multimodal Framework Overview}
SPAR-Net was designed with two key objectives: (1) to develop a scalable and accessible approach to real-time image guidance using standard clinical equipment, and (2) to achieve computationally efficient motion estimation suitable for real-time applications ($<$500 ms). 
SPAR-Net adopts a dual-branch, U-shaped encoder–decoder architecture designed to integrate anatomical and physiological information. The imaging branch employs a frozen RAD-DINO encoder to extract multiscale anatomical representations from X-ray projections, with an optional domain adaptation module incorporated for cardiac segmentation to address imaging domain shifts. The physiological branch uses a Mamba-based encoder to model temporal dependencies in ECG and respiratory signals.
Physiological information is integrated through the MPSRU and the CRU. The MPSRU identifies anatomical regions where physiological information is most relevant, while the CRU adaptively regulates the contribution of physiological features across image channels, enabling selective multimodal fusion. Physiological guidance is applied to intermediate RAD-DINO features from layers 3 and 6, while features from layers 9 and 12 are passed directly to the decoder to preserve higher-level anatomical representations. The resulting multiscale features are progressively combined through a U-Net decoder with skip connections and residual dilated convolutions, enabling the recovery of spatial detail and the delineation of target boundaries. A final 1 × 1 convolution produces binary segmentation logits at the original image resolution. The overall architecture is illustrated in Fig.~\ref{fig:jbhi_fig2}.

\subsection{Image and Physiological Feature Encoding}
\subsubsection{Multi-Level Image Feature Extraction and Gated Domain Adaptation}

To extract complementary image representations, the proposed framework employs a pretrained RAD-DINO encoder for semantic feature extraction, with an optional SimSiam-pretrained ResNet-18 domain encoder to facilitate adaptation between DRR and X-ray domains.

Given an input image $\mathbf{I}_i \in \mathbb{R}^{B \times H \times W \times 1}$, RAD-DINO extracts feature representations from Transformer layers $\ell \in \{3,6,9,12\}$. The patch tokens from each selected layer are rearranged into spatial feature maps:

\begin{equation}
\mathbf{R}_{\ell}
=
E_{\mathrm{RAD}}^{(\ell)}(\mathbf{I}_i),
\quad
\ell \in \{3,6,9,12\}.
\label{eq:rad_dino_features}
\end{equation}

where $E_{\mathrm{RAD}}^{(\ell)}$ denotes feature extraction from the $\ell$-th RAD-DINO transformer layer, and $\mathbf{R}_{\ell} \in \mathbb{R}^{B \times H_P \times W_P \times C}$ represents the corresponding spatial feature map, with $H_P$ and $W_P$ denoting the spatial dimensions of the patch feature map.

For the cardiac task, the SimSiam-pretrained ResNet-18 encoder additionally extracts hierarchical domain-specific features from four convolutional stages:

\begin{equation}
\mathbf{D}_{\ell}
=
E_{\mathrm{dom}}^{(\ell)},
\quad
\ell \in \{3,6,9,12\}.
\label{eq:domain_encoder_features}
\end{equation}

Here, $\mathbf{D}_{\ell}$ represents the ResNet-18 feature map paired with RAD-DINO layer $\ell$, with the four convolutional stages aligned in order with RAD-DINO layers 3, 6, 9, and 12.

To address the DRR-to-X-ray domain shift, a gated residual feature fusion mechanism is introduced. At each selected level, the ResNet-18 features are spatially aligned with the corresponding RAD-DINO representations through bilinear interpolation and projected to a compatible feature dimension. The aligned representations are concatenated and processed through two parallel convolutional branches to generate a residual feature adjustment $\Delta\mathbf{R}_{\ell}$ and a spatial gating map $\mathbf{G}_{\ell}$, respectively. The gating map is constrained to $(0,1)$ through sigmoid activation, enabling spatially selective modulation of the residual features.

The adapted representation is obtained as:

\begin{equation}
\mathbf{F}_{\ell}
=
\mathbf{R}_{\ell}
+
\alpha_{\ell}
\left(
\mathbf{G}_{\ell}
\odot
\Delta\mathbf{R}_{\ell}
\right).
\label{eq:gated_residual_fusion}
\end{equation}

where $\odot$ denotes element-wise multiplication with broadcasting across feature channels, and $\alpha_{\ell}$ is a learnable, bounded scaling coefficient controlling the contribution of domain-adaptive residual features. The final convolution of the residual branch is zero-initialized, ensuring that the fusion initially preserves the pretrained RAD-DINO representations.

For the lung and liver tasks, the domain encoder is disabled, and the original RAD-DINO features are directly retained as $\mathbf{F}_{\ell}=\mathbf{R}_{\ell}$. The resulting multi-level representations $\mathbf{F}_{\ell}$ are subsequently used for physiological guidance and segmentation decoding in both tasks.

\subsubsection{Causal Physiological Sequence Encoding}

Physiological signals provide complementary temporal information for anatomical motion modeling. Given a continuous physiological signal $\mathbf{S} \in \mathbb{R}^{L \times 1}$, where $L$ denotes the total number of signal samples, a fixed-length causal window $\mathbf{W}_i \in \mathbb{R}^{L_w \times 1}$ is constructed for each input image based on its temporally aligned signal index $t_i$:

\begin{equation}
\mathbf{W}_i =
\mathbf{S}
\left[
\max(0, t_i - L_w + 1) : t_i + 1
\right].
\label{eq:causal_window}
\end{equation}

Here, $L_w$ denotes the predefined window length. Each window contains the current signal measurement and its preceding samples, excluding future information. When insufficient historical samples are available, the window is left-padded by repeating the first signal value. Each sample within the window is independently projected into a $d$-dimensional feature space using a shared learnable linear embedding layer, followed by layer normalization:

\begin{equation}
\mathbf{Z}_i =
\mathrm{LN}_{\mathrm{in}}
\left(
\mathbf{W}_i \mathbf{W}_s + \mathbf{b}_s
\right).
\label{eq:physiological_embedding}
\end{equation}

where $\mathbf{W}_s \in \mathbb{R}^{1 \times d}$ and $\mathbf{b}_s \in \mathbb{R}^{d}$ are learnable projection parameters, and $\mathbf{Z}_i \in \mathbb{R}^{L_w \times d}$ denotes the embedded physiological sequence.

The embedded sequence is subsequently processed by a Mamba encoder $E_{\mathrm{mamba}}$ to capture temporal dependencies:

\begin{equation}
\mathbf{H}_i =
\operatorname{Dropout}
\left[
\mathrm{LN}_{\mathrm{out}}
\left(
E_{\mathrm{mamba}}(\mathbf{Z}_i)
\right)
\right].
\label{eq:mamba_encoding}
\end{equation}

From the Mamba output features $\mathbf{H}_i \in \mathbb{R}^{L_w \times d}$, the $k$ most recent hidden states $\mathbf{H}_i^{k} = \mathbf{H}_i[L_w-k:L_w, :] \in \mathbb{R}^{k \times d}$ are selected to provide temporally relevant physiological representations for cross-modal attention while incorporating historical context encoded by Mamba. The current hidden state $\mathbf{H}_i^{\mathrm{cur}} = \mathbf{H}_i[L_w-1, :] \in \mathbb{R}^{d}$ is additionally used to provide physiological context for channel relevance estimation. The physiological encoder is jointly optimized with the downstream segmentation network.

\subsection{Masked Physiology-guided Spatial Relevance Unit (MPSRU)}
Although physiological signals provide temporal information about
anatomical motion, their relevance to image features varies spatially,
as image appearance is also influenced by gantry angle dependent
projection changes and overlapping structures. To distinguish
physiologically relevant regions from unrelated image variations,
we introduce the MPSRU, which learns spatial relevance through
masked cross-modal attention and constructs physiology-conditioned
image representations.

The MPSRU is independently applied to RAD-DINO layers
$\ell \in \{3,6\}$. For each selected layer, the image feature
map $\mathbf{F}_{\ell}$ is flattened into
$N = H_P W_P$ spatial tokens:

\begin{equation}
\mathbf{X}_{\ell}
=
\operatorname{Flatten}(\mathbf{F}_{\ell})
\in \mathbb{R}^{B \times N \times C}.
\label{eq:mpsru_flatten}
\end{equation}

The selected Mamba hidden states $\mathbf{H}_i^{k}$ serve as
physiological queries, while the image patch tokens
$\mathbf{X}_{\ell}$ provide keys and values for cross-modal
attention. During training, each spatial patch token is
independently masked with probability $\rho$, using a binary
mask $\mathbf{M}_{\ell} \in \{0,1\}^{B \times N}$:

\begin{equation}
\begin{aligned}
P(M_{\ell,p}=0) &= \rho,\\
P(M_{\ell,p}=1) &= 1-\rho.
\end{aligned}
\label{eq:mpsru_mask_probability}
\end{equation}

Here, $M_{\ell,p}=0$ indicates that the $p$-th image feature
patch token is excluded from attention, while $M_{\ell,p}=1$
indicates that it is retained. Each masking decision applies
to the entire patch feature vector across all channels and
is shared across attention heads and physiological queries.
Importantly, masking is applied only to the cross-modal
attention computation, leaving the original image feature
map unchanged. All spatial patch tokens are retained during
inference.

The masked cross-modal attention
$\operatorname{CA}_{\mathrm{mask}}$ is formulated as:

\begin{equation}
(\mathbf{A}_{\ell},\mathbf{O}_{\ell})
=
\operatorname{CA}_{\mathrm{mask}}
\left(
\mathbf{H}_i^{k},
\mathbf{X}_{\ell};
\mathbf{M}_{\ell}
\right).
\label{eq:mpsru_cross_attention}
\end{equation}

where $\mathbf{A}_{\ell}$ and $\mathbf{O}_{\ell}$
denote the cross-modal attention weights and attended
image context, respectively.

To estimate spatial relevance, the attention weights are
processed through a Spatial Attention Mapping operation,
which averages the attention weights across heads and
physiological queries, rescales them according to the number
of valid tokens, and clips the resulting values. The
aggregated attention is reshaped to the patch grid and
processed by a lightweight convolutional network:

\begin{equation}
\mathbf{S}_{\ell}
=
f_{\mathrm{spatial},\ell}
\left(
\operatorname{Mean}_{\mathrm{heads,queries}}
(\mathbf{A}_{\ell})
\right).
\label{eq:mpsru_spatial_relevance}
\end{equation}

The function $f_{\mathrm{spatial},\ell}$ incorporates
valid-token rescaling, clipping, spatial reshaping,
convolutional refinement, and sigmoid activation.
The resulting
$\mathbf{S}_{\ell}
\in (0,1)^{B \times 1 \times H_P \times W_P}$
represents a learned spatial relevance map rather than
an explicitly supervised anatomical motion field.

In parallel, the attended image context undergoes
Physiological Context Projection, where the representations
are averaged across the $k$ physiological queries and
linearly projected into the image feature dimension to
generate a context vector
$\mathbf{c}_{\ell} \in \mathbb{R}^{B \times C}$:

\begin{equation}
\mathbf{c}_{\ell}
=
f_{\mathrm{context},\ell}
\left(
\operatorname{Mean}_{k}
(\mathbf{O}_{\ell})
\right).
\label{eq:mpsru_context_projection}
\end{equation}

Here, $f_{\mathrm{context},\ell}$ denotes a learnable
linear projection.

The context vector $\mathbf{c}_{\ell}$ is then spatially
expanded and concatenated with the projected image features
along the channel dimension. The combined features are
processed by a lightweight convolutional fusion module
to generate the physiology-conditioned representation:

\begin{equation}
\mathbf{P}_{\ell}
=
f_{\mathrm{phy},\ell}
\left(
\operatorname{Concat}
\left(
f_{\mathrm{img},\ell}(\mathbf{F}_{\ell}),
\mathbf{c}_{\ell}
\right)
\right).
\label{eq:mpsru_physiology_conditioned}
\end{equation}

Here, $f_{\mathrm{img},\ell}$ denotes a
$1 \times 1$ convolutional projection, while
$f_{\mathrm{phy},\ell}$ consists of a
$1 \times 1$ convolution followed by group normalization,
ReLU activation, and a $3 \times 3$ convolution.
The resulting
$\mathbf{P}_{\ell}
\in \mathbb{R}^{B \times C \times H_P \times W_P}$,
together with $\mathbf{S}_{\ell}$, is subsequently used
for spatial--channel relevance guided residual fusion.

\subsection{Channel Relevance Unit (CRU)}
While the MPSRU identifies spatial regions associated with
physiological information, the relevance of physiological
guidance may also vary across image feature channels.
To account for this variation, we introduce the CRU,
which adaptively estimates channel-wise relevance by
integrating image features, physiology-conditioned
representations, and the current physiological state.

Specifically, the image feature map $\mathbf{F}_{\ell}$
and physiology-conditioned feature map $\mathbf{P}_{\ell}$
are independently aggregated through global average
pooling (GAP). The resulting feature vectors are
concatenated with the final Mamba hidden state
$\mathbf{H}_i^{\mathrm{cur}}$ and passed through
a lightweight channel gating network:

\begin{equation}
\begin{aligned}
\mathbf{C}_{\ell}
= f_{\mathrm{channel},\ell}\big(
&\operatorname{Concat}(
\operatorname{GAP}(\mathbf{F}_{\ell}),\\
&\operatorname{GAP}(\mathbf{P}_{\ell}),
\mathbf{H}_i^{\mathrm{cur}})
\big).
\end{aligned}
\label{eq:channel_relevance}
\end{equation}

where $f_{\mathrm{channel},\ell}$ consists of two
fully connected layers with an intermediate ReLU
activation and a final sigmoid function. The resulting
$\mathbf{C}_{\ell}
\in (0,1)^{B \times C \times 1 \times 1}$
represents the channel-wise relevance weights.

\subsection{Adaptive Spatial–Channel Residual Fusion}
The spatial relevance map $\mathbf{S}_{\ell}$ from MPSRU
and channel relevance weights $\mathbf{C}_{\ell}$ from CRU
are jointly applied to the physiology-conditioned feature
map $\mathbf{P}_{\ell}$. The resulting features are
incorporated into the original image representation
through a learnable residual connection:

\begin{equation}
\mathbf{F}_{\ell}^{\mathrm{guided}}
=
\mathbf{F}_{\ell}
+
\beta_{\ell}
\left(
\mathbf{P}_{\ell}
\odot
\mathbf{S}_{\ell}
\odot
\mathbf{C}_{\ell}
\right).
\label{eq:physiological_guidance_fusion}
\end{equation}

where $\odot$ denotes element-wise multiplication
with broadcasting across spatial and channel dimensions,
and $\beta_{\ell}$ is a learnable scalar controlling
the contribution of physiological guidance.

This formulation preserves the original image features
while selectively incorporating physiological information
according to its spatial and channel relevance.
The resulting $\mathbf{F}_{\ell}^{\mathrm{guided}}$
is subsequently passed to the segmentation decoder.

\subsection{Segmentation Prediction and Optimization Objective}
The proposed MPSRU and CRU are designed as modular components
that can be flexibly integrated at different levels of
encoder--decoder architectures. Their placement can be
adjusted according to the task requirements and feature
characteristics, allowing physiological guidance to be
incorporated at selected feature levels.

The decoder produces a single-channel segmentation logit
map at the original image resolution. A sigmoid activation
converts the logits into pixelwise foreground probabilities.
The segmentation network is optimized using a combination
of binary cross-entropy (BCE) and Dice losses:

\begin{equation}
\mathcal{L}_{\mathrm{seg}}
=
\lambda_{\mathrm{BCE}}
\mathcal{L}_{\mathrm{BCE}}
+
\lambda_{\mathrm{Dice}}
\mathcal{L}_{\mathrm{Dice}}.
\label{eq:segmentation_loss}
\end{equation}

where $\lambda_{\mathrm{BCE}}$ and
$\lambda_{\mathrm{Dice}}$ control the contributions
of the two losses, both set to 1 in our experiments.
The pretrained RAD-DINO encoder remains frozen, while
the trainable components are jointly optimized using
the segmentation objective.

\section{Experimental Design and Evaluation Methods}
\label{sec:Experimental Design and Evaluation Methods}
\subsection{Implementation Details}
For the cardiac dataset, subject-specific training strategy
was adopted for each sheep. Data augmentation was performed by randomly
translating the planning CT geometry by up to 10~mm and rotating it by
up to $10^\circ$ for generating additional DRRs~\cite{ref15}. This process was
repeated 100 times to simulate treatment setup variations and anatomical
motion. The resulting DRR dataset was divided into training and
validation sets at a ratio of 80\% to 20\%, while real X-ray projections
from a separate cone-beam computed tomography (CBCT) acquisition were used for testing. The framework employed a ResNet-18 encoder trained from scratch. Training was performed for up to 100 epochs with a batch size of 24, using a symmetric negative cosine similarity loss with stop-gradient
operations. Stochastic gradient descent (SGD) was employed with a
momentum of 0.9 and weight decay of 0.0001. Cosine learning rate decay
was applied to the encoder and projection head, while the prediction
head maintained a fixed learning rate. 

SPAR-Net employed a frozen RAD-DINO encoder and a SimSiam-pretrained ResNet-18 domain encoder for multi-level feature extraction and adaptation. ECG signals were temporally synchronized with the corresponding X-ray projections and processed using a single-layer Mamba network with a hidden dimension of 32. The ECG window length was set to 80 samples (approximately 0.46 s) to capture recent cardiac dynamics. The 4 most recent Mamba hidden states were selected for cross-modal attention to emphasize physiological information close to the current X-ray acquisition while preserving temporal context. The embedding dimension was set to 32, with a single Mamba layer. Physiological guidance was integrated through four-head cross-attention with an embedding dimension of 96, incorporating spatial and channel relevance weighting. A spatial masking ratio of 25\% was applied during training, while the decoder employed multi-scale dilated convolutions with dilation rates of 1, 2, 4, and 6. The framework was trained for up to 10 epochs using the AdamW optimizer with a weight decay of 0.0001. Initial learning rates were set to 0.0002 for the decoder and physiological guidance modules and 0.0001 for the ECG encoder and domain adaptation module. A combined binary cross-entropy and Dice loss was employed with equal weighting.

For the lung and liver dataset, As both training and testing
involved X-ray images, the SimSiam-pretrained ResNet-18 domain encoder
was excluded. Respiratory signals were temporally synchronized with the
corresponding X-ray projections and processed using the same
Mamba-based approach as the ECG signals in the cardiac setting.
The respiratory surface motion signals window length was set to
32 samples to capture recent respiratory dynamics. The four most
recent Mamba hidden states were selected for cross-modal attention. All input images were resized to $512 \times 512$ pixels. Data augmentation included random horizontal flipping, mild affine transformations with rotations of $\pm 3^\circ$, translations of
$\pm 3\%$, and scaling between 0.97 and 1.03, together with random
brightness and contrast adjustments and Gaussian noise. The segmentation framework followed the same configuration as that used for the cardiac dataset, except that training was extended to a maximum of 100 epochs.

All experiments were conducted on two NVIDIA RTX PRO 6000 Blackwell GPU.

\subsection{Evaluation Setup}
To evaluate segmentation performance, we employed the Dice similarity coefficient (Dice), 95th percentile Hausdorff distance (HD95), mean surface distance (MSD), and centroid distance (CD). Dice measures the spatial overlap between predicted and ground-truth (GT) segmentation masks, while HD95 and MSD quantify boundary discrepancies based on bidirectional surface distances. Specifically, HD95 represents the 95th percentile of the combined surface distances, whereas MSD measures their mean. CD evaluates spatial localization accuracy by calculating the Euclidean distance between the centroids of the largest external contours of the predicted and GT masks. An in-plane pixel spacing of 0.256 mm × 0.256 mm was used for all distance calculations, with HD95, MSD, and CD reported in millimeters (mm).

\subsection{Comparison with State-of-the-Art Medical Image Segmentation Methods}

We compared SPAR-Net with established medical image segmentation methods, including (i) U-Net, a convolutional encoder–decoder architecture with skip connections for multi-scale feature integration; (ii) DeepLabV3+, which employs atrous convolutions to capture multi-scale spatial context; (iii) TransUNet, which combines convolutional feature extraction with transformer-based global context modeling; (iv) U-Mamba, which incorporates state-space modeling to capture long-range dependencies; and (v) MedSAM, a foundation model adapted for medical image segmentation. For MedSAM, the pretrained image encoder was frozen, while the remaining trainable components were fine-tuned using our dataset. MedSAM was evaluated only on the animal cardiac and lung datasets, as liver tumors are not directly visible in X-ray projections, making image-based prompting unsuitable for the liver segmentation task. These comparisons allowed us to assess the effectiveness of SPAR-Net against convolutional, transformer-based, state-space, and foundation model-based approaches.

For all methods, the predicted segmentation masks underwent the same post-processing procedures, including hole filling and removal of small disconnected regions. As each image contained a single target structure, only the largest connected component was retained to eliminate spurious predictions and ensure consistent evaluation.

\subsection{Ablation study}
We conducted ablation studies on the cardiac dataset to investigate the contributions of individual components of SPAR-Net. First, an image-only model using the frozen RAD-DINO encoder was established as the baseline. Next, the domain encoder was incorporated to evaluate its contribution to addressing imaging domain shifts. Subsequently, the MPSRU and CRU were introduced to integrate physiological information, while random feature masking within the MPSRU was disabled. Finally, random feature masking was enabled to obtain the full SPAR-Net and evaluate its additional contribution to segmentation performance.

All comparison methods were evaluated using the same training and testing protocols to ensure a fair comparison.


\begin{table*}[!t]
\centering
\caption{Comparison of SPAR-Net with state-of-the-art segmentation methods on the cardiac dataset across three test animals, including inference time. Results are reported as the mean $\pm$ standard deviation of image-level HD95, MSD, CD, and Dice values for each animal. Overall results are calculated by averaging the animal-level means, with each animal contributing equally. Inference time is reported in milliseconds for processing a single test image. The best scores are highlighted in bold, and the second-best scores are underlined.}
\label{tab:cardiac_results}

\setlength{\tabcolsep}{3pt}
\renewcommand{\arraystretch}{1.9}
\resizebox{\textwidth}{!}{%
\begin{tabular}{l*{16}{c}c}
\toprule
\multirow{2}{*}{Methods}
& \multicolumn{4}{c}{Animal 1}
& \multicolumn{4}{c}{Animal 2}
& \multicolumn{4}{c}{Animal 3}
& \multicolumn{4}{c}{Overall}
& \multirow{2}{*}{\shortstack{Inference\\Time}} \\
\cmidrule(lr){2-5}
\cmidrule(lr){6-9}
\cmidrule(lr){10-13}
\cmidrule(lr){14-17}

& HD95 & MSD & CD & Dice
& HD95 & MSD & CD & Dice
& HD95 & MSD & CD & Dice
& HD95 & MSD & CD & Dice
& (ms/image) \\
\midrule

UNet
& $23.8\pm11.5$ & $8.0\pm4.3$ & $7.1\pm4.9$ & $0.82\pm0.09$
& $28.1\pm17.9$ & $9.5\pm5.8$ & $10.9\pm11.0$ & $0.78\pm0.16$
& $15.6\pm10.5$ & $3.5\pm2.0$ & $5.3\pm4.2$ & $0.93\pm0.04$
& $22.5\pm6.4$ & $7.0\pm3.1$ & $7.8\pm2.8$ & $0.85\pm0.08$
& $\mathbf{29}$ \\

DeepLabV3
& $\underline{13.1\pm8.2}$ & $\mathbf{4.9\pm2.1}$ & $\mathbf{5.8\pm2.6}$ & $\underline{0.89\pm0.04}$
& $23.7\pm6.9$ & $8.5\pm2.9$ & $8.8\pm4.8$ & $0.81\pm0.09$
& $8.9\pm7.4$ & $\underline{2.3\pm1.3}$ & $\underline{2.9\pm2.1}$ & $\underline{0.95\pm0.02}$
& $15.2\pm7.6$ & $5.2\pm3.1$ & $\underline{5.8\pm3.0}$ & $0.88\pm0.07$
& $\underline{34}$ \\

TransUNet
& $\underline{13.1\pm9.1}$ & $\underline{5.0\pm2.7}$ & $\underline{6.2\pm3.5}$ & $\underline{0.89\pm0.05}$
& $\underline{16.5\pm10.7}$ & $\underline{5.8\pm2.2}$ & $\underline{8.5\pm5.3}$ & $\underline{0.89\pm0.03}$
& $14.2\pm8.0$ & $3.6\pm1.6$ & $6.2\pm3.8$ & $0.92\pm0.03$
& $\underline{14.6\pm1.8}$ & $\underline{4.8\pm1.2}$ & $7.0\pm1.4$ & $\underline{0.90\pm0.02}$
& $144$ \\

UMamba
& $34.1\pm27.8$ & $15.35\pm15.31$ & $27.68\pm27.15$ & $0.66\pm0.26$
& $35.0\pm32$ & $14.0\pm12.7$ & $24.5\pm26.0$ & $0.70\pm0.27$
& $26.4\pm20.4$ & $9.1\pm7.8$ & $18.1\pm17.0$ & $0.78\pm0.18$
& $32.0\pm4.9$ & $12.8\pm3.3$ & $23.4\pm4.9$ & $0.71\pm0.06$
& $66$ \\

MedSAM
& $17.4\pm4.4$ & $7.5\pm1.0$ & $9.4\pm1.5$ & $0.83\pm0.03$
& $23.5\pm8.9$ & $9.6\pm2.8$ & $10.9\pm3.8$ & $0.81\pm0.05$
& $\underline{6.8\pm1.3}$ & $\underline{2.3\pm0.5}$ & $3.6\pm0.9$ & $0.94\pm0.01$
& $15.9\pm8.4$ & $6.5\pm3.8$ & $7.9\pm3.8$ & $0.86\pm0.07$
& $80$ \\

SPAR-Net
& $\mathbf{11.0\pm5.1}$ & $\mathbf{4.9\pm2.0}$ & $6.8\pm4.2$ & $\mathbf{0.90\pm0.04}$
& $\mathbf{11.1\pm3.7}$ & $\mathbf{5.0\pm2.4}$ & $\mathbf{7.4\pm4.9}$ & $\mathbf{0.90\pm0.05}$
& $\mathbf{5.1\pm3.0}$ & $\mathbf{1.8\pm0.9}$ & $\mathbf{2.7\pm1.8}$ & $\mathbf{0.96\pm0.02}$
& $\mathbf{8.7\pm3.3}$ & $\mathbf{3.8\pm1.8}$ & $\mathbf{5.3\pm2.3}$ & $\mathbf{0.92\pm0.04}$
& $116$ \\

\bottomrule
\end{tabular}%
}
\end{table*}


\begin{table*}[!t]
\centering
\caption{Comparison of SPAR-Net with state-of-the-art segmentation methods on the lung tumor segmentation dataset across two test patients. Results are reported as the mean $\pm$ standard deviation of image-level HD95, MSD, CD, and Dice values for each patient. Overall results are calculated by averaging the patient-level means, with each patient contributing equally. The best scores are highlighted in bold, and the second-best scores are underlined.}
\label{tab:lung_results}

\setlength{\tabcolsep}{3pt}
\renewcommand{\arraystretch}{1.5}

\resizebox{\textwidth}{!}{%
\begin{tabular}{l*{12}{c}}
\toprule
\multirow{2}{*}{Methods}
& \multicolumn{4}{c}{PAT01}
& \multicolumn{4}{c}{PAT02}
& \multicolumn{4}{c}{Overall} \\
\cmidrule(lr){2-5}
\cmidrule(lr){6-9}
\cmidrule(lr){10-13}
& HD95 & MSD & CD & Dice
& HD95 & MSD & CD & Dice
& HD95 & MSD & CD & Dice \\
\midrule

UNet
& $3.8\pm4.7$ & $1.7\pm3.5$ & $2.1\pm3.2$ & $0.76\pm0.24$
& $3.3\pm2.0$ & $\underline{1.2\pm0.7}$ & $\mathbf{1.5\pm1.1}$ & $0.91\pm0.05$
& $3.5\pm0.48$ & $1.5\pm0.4$ & $\underline{1.7\pm0.5}$ & $0.84\pm0.11$ \\

DeepLabV3
& $\underline{3.0\pm1.2}$ & $\underline{1.3\pm0.6}$ & $\underline{1.9\pm1.2}$ & $0.84\pm0.07$
& $\underline{3.0\pm1.1}$ & $\underline{1.2\pm0.5}$ & $\underline{1.7\pm0.9}$ & $\underline{0.92\pm0.03}$
& $\underline{3.0\pm0.01}$ & $\underline{1.2\pm0.05}$ & $\underline{1.7\pm0.1}$ & $0.88\pm0.06$ \\

TransUNet
& $3.3\pm4.7$ & $1.6\pm3.5$ & $2.3\pm4.5$ & $0.83\pm0.15$
& $3.2\pm1.1$ & $1.3\pm0.5$ & $1.8\pm1.1$ & $0.91\pm0.03$
& $3.2\pm0.07$ & $1.5\pm0.2$ & $2.0\pm0.3$ & $0.87\pm0.05$ \\

UMamba
& $4.9\pm11.7$ & $3.1\pm10.1$ & $4.1\pm11.8$ & $0.81\pm0.20$
& $3.4\pm1.6$ & $1.3\pm0.6$ & $1.9\pm1.4$ & $0.91\pm0.04$
& $4.1\pm1.1$ & $2.2\pm1.3$ & $3.0\pm1.6$ & $0.86\pm0.07$ \\

MedSAM
& $3.2\pm1.2$ & $1.4\pm0.62$ & $2.1\pm1.3$ & $\mathbf{0.87\pm0.05}$
& $3.2\pm1.1$ & $1.3\pm0.5$ & $1.8\pm1.1$ & $\mathbf{0.94\pm0.02}$
& $3.2\pm0.02$ & $1.4\pm0.1$ & $1.9\pm0.3$ & $\mathbf{0.91\pm0.04}$ \\

SPAR-Net
& $\mathbf{2.5\pm1.0}$ & $\mathbf{1.1\pm0.5}$ & $\mathbf{1.6\pm1.0}$ & $\underline{0.85\pm0.07}$
& $\mathbf{2.8\pm1.3}$ & $\mathbf{1.0\pm0.5}$ & $\mathbf{1.5\pm1.1}$ & $\underline{0.92\pm0.03}$
& $\mathbf{2.6\pm0.2}$ & $\mathbf{1.0\pm0.03}$ & $\mathbf{1.5\pm0.1}$ & $\underline{0.89\pm0.05}$ \\

\bottomrule
\end{tabular}%
}
\end{table*}


\begin{table*}[!t]
\centering
\caption{Comparison of SPAR-Net with state-of-the-art segmentation methods on the liver tumor segmentation dataset across five test patients. Results are reported as the mean $\pm$ standard deviation of image-level HD95, MSD, CD, and Dice values for each patient. Overall results are calculated by averaging the patient-level means, with each patient contributing equally. The best scores are highlighted in bold, and the second-best scores are underlined.}
\label{tab:liver_results}

\setlength{\tabcolsep}{3pt}
\renewcommand{\arraystretch}{2.0}

\resizebox{\textwidth}{!}{%
\begin{tabular}{l*{24}{c}}
\toprule
\multirow{2}{*}{Methods}
& \multicolumn{4}{c}{PAT01}
& \multicolumn{4}{c}{PAT02}
& \multicolumn{4}{c}{PAT03}
& \multicolumn{4}{c}{PAT04}
& \multicolumn{4}{c}{PAT05}
& \multicolumn{4}{c}{Overall} \\
\cmidrule(lr){2-5}
\cmidrule(lr){6-9}
\cmidrule(lr){10-13}
\cmidrule(lr){14-17}
\cmidrule(lr){18-21}
\cmidrule(lr){22-25}

& HD95 & MSD & CD & Dice
& HD95 & MSD & CD & Dice
& HD95 & MSD & CD & Dice
& HD95 & MSD & CD & Dice
& HD95 & MSD & CD & Dice
& HD95 & MSD & CD & Dice \\
\midrule

UNet
& $11.7\pm9.4$ & $4.4\pm3.7$ & $6.1\pm6.3$ & $0.78\pm0.21$
& $4.0\pm2.8$ & $1.2\pm0.6$ & $\underline{1.4\pm1.0}$ & $0.95\pm0.02$
& $11.2\pm5.1$ & $4.2\pm1.8$ & $5.7\pm3.1$ & $0.77\pm0.11$
& $6.1\pm4.0$ & $2.2\pm1.4$ & $3.0\pm2.5$ & $0.89\pm0.08$
& $14.7\pm4.8$ & $5.0\pm1.7$ & $6.3\pm3.3$ & $0.81\pm0.07$
& $9.5\pm4.4$ & $3.4\pm1.6$ & $4.5\pm2.2$ & $0.84\pm0.08$ \\

DeepLabV3
& $4.0\pm1.2$ & $\mathbf{1.7\pm0.6}$ & $\mathbf{2.6\pm1.2}$ & $\mathbf{0.93\pm0.02}$
& $\underline{2.5\pm1.2}$ & $\underline{1.0\pm0.4}$ & $1.5\pm0.7$ & $\underline{0.96\pm0.01}$
& $6.0\pm2.2$ & $2.6\pm1.1$ & $3.5\pm2.0$ & $0.88\pm0.05$
& $2.4\pm1.1$ & $\underline{1.0\pm0.4}$ & $\mathbf{1.3\pm0.8}$ & $\mathbf{0.95\pm0.02}$
& $4.2\pm1.5$ & $1.8\pm0.7$ & $2.8\pm1.4$ & $0.93\pm0.02$
& $3.8\pm1.5$ & $\underline{1.6\pm0.7}$ & $\underline{2.3\pm0.9}$ & $\underline{0.93\pm0.03}$ \\

TransUNet
& $5.0\pm2.0$ & $2.0\pm0.8$ & $3.0\pm1.5$ & $0.91\pm0.03$
& $2.6\pm1.0$ & $\underline{1.0\pm0.3}$ & $\underline{1.4\pm0.8}$ & $\underline{0.96\pm0.01}$
& $\underline{5.2\pm2.0}$ & $\underline{2.2\pm0.9}$ & $\underline{3.1\pm1.6}$ & $\underline{0.89\pm0.04}$
& $3.0\pm1.7$ & $1.2\pm0.6$ & $1.6\pm1.0$ & $0.94\pm0.03$
& $4.5\pm2.0$ & $1.8\pm0.8$ & $\underline{2.3\pm1.5}$ & $0.93\pm0.02$
& $4.0\pm1.2$ & $\underline{1.6\pm0.5}$ & $\underline{2.3\pm0.8}$ & $\underline{0.93\pm0.03}$ \\

UMamba
& $\underline{3.90\pm1.42}$ & $1.83\pm0.75$ & $2.90\pm1.41$ & $\mathbf{0.93\pm0.03}$
& $2.7\pm1.5$ & $1.15\pm0.50$ & $1.88\pm0.89$ & $0.95\pm0.02$
& $5.4\pm2.6$ & $2.5\pm1.4$ & $3.9\pm2.5$ & $0.88\pm0.06$
& $\underline{2.3\pm1.0}$ & $1.1\pm0.6$ & $1.8\pm1.0$ & $\mathbf{0.95\pm0.03}$
& $\underline{3.93\pm1.6}$ & $\underline{1.70\pm0.77}$ & $2.4\pm1.5$ & $\underline{0.94\pm0.03}$
& $\underline{3.4\pm1.2}$ & $\underline{1.6\pm0.6}$ & $2.5\pm0.9$ & $\underline{0.93\pm0.03}$ \\

MedSAM
& \multicolumn{24}{c}{Not applicable to invisible targets} \\

SPAR-Net
& $\mathbf{3.7\pm1.4}$ & $\mathbf{1.7\pm0.7}$ & $\mathbf{2.6\pm1.4}$ & $\mathbf{0.93\pm0.03}$
& $\mathbf{2.0\pm0.6}$ & $\mathbf{0.8\pm0.2}$ & $\mathbf{1.1\pm0.5}$ & $\mathbf{0.97\pm0.01}$
& $\mathbf{4.0\pm1.5}$ & $\mathbf{1.8\pm0.8}$ & $\mathbf{2.8\pm1.4}$ & $\mathbf{0.91\pm0.04}$
& $\mathbf{2.2\pm0.8}$ & $\mathbf{0.9\pm0.4}$ & $\mathbf{1.3\pm0.6}$ & $\mathbf{0.95\pm0.01}$
& $\mathbf{3.6\pm1.4}$ & $\mathbf{1.5\pm0.7}$ & $\mathbf{2.0\pm1.4}$ & $\mathbf{0.95\pm0.02}$
& $\mathbf{3.2\pm0.9}$ & $\mathbf{1.4\pm0.4}$ & $\mathbf{2.0\pm0.7}$ & $\mathbf{0.94\pm0.02}$ \\

\bottomrule
\end{tabular}%
}
\end{table*}

\section{Results}
In this section, we present the performance of SPAR-Net across three anatomical sites. Tables~\ref{tab:cardiac_results}--\ref{tab:liver_results} present the cardiac, lung tumor and liver tumor segmentation results, respectively, including comparisons between SPAR-Net and state-of-the-art segmentation methods. Table~\ref{tab:ablation} presents an ablation study evaluating the contributions of the domain encoder, physiological signals, and integration modules to segmentation performance. Fig.~\ref{fig:jbhi_fig3} presents the quantitative segmentation results, while Figs.~\ref{fig:jbhi_fig4}--\ref{fig:jbhi_fig6} provide representative segmentation examples across the three anatomical sites, illustrating the improved boundary delineation achieved by SPAR-Net compared with competing methods.

\subsection{Performance Comparison on the Cardiac Dataset}
The cardiac segmentation results are presented in Table~\ref{tab:cardiac_results}. 
SPAR-Net achieved the best overall performance across all four metrics, with an HD95 of 8.7 mm, MSD of 3.8 mm, CD of 5.3 mm, and Dice of 0.92. Compared with TransUNet, which achieved the second-best overall HD95 of 14.6 mm, SPAR-Net reduced HD95 by 40.4\ and MSD by 20.8\%, while improving Dice from 0.90 to 0.92. DeepLabV3 achieved the second-best overall CD of 5.8 mm, compared with 5.3 mm for SPAR-Net. Regarding computational efficiency, SPAR-Net required 116 ms per X-ray image, faster than TransUNet (144 ms), while remaining within the predefined of 500 ms per image meeting real-time image-guidance criterion~\cite{ref50}.

At the individual animal level, SPAR-Net achieved the lowest HD95 across all three animals. The advantage was particularly evident for Animal 2, where HD95 decreased by 32.7\% compared with the strongest baseline. As shown in the violin plots in Fig.~\ref{fig:jbhi_fig3}a, SPAR-Net also exhibited more concentrated distributions across HD95, MSD, CD, and Dice, whereas several baseline methods showed greater image-level variation, particularly in the distance-based metrics. 

Representative segmentation results are presented in Fig. ~\ref{fig:jbhi_fig4}. Under low-contrast imaging conditions, UNet produced inaccurate segmentation masks with uneven boundaries and missing regions, while UMamba exhibited noticeable contour distortions and incomplete segmentation regions. DeepLabV3 and TransUNet generated more accurate masks, although some boundary errors remained. MedSAM performed less accurately than these two methods, tending to undersegment the cardiac region, particularly in low-contrast areas. In contrast, SPAR-Net produced continuous masks with smoother contours and shapes closely matching the GT. Overall, SPAR-Net achieved more consistent boundary delineation and target localization across the three animals, particularly in cases with greater training-to-testing domain shifts.

\subsection{Performance Comparison on the Lung Tumor Dataset}
\begin{figure*}[!t]
    \centering
    \includegraphics[width=0.8\textwidth]{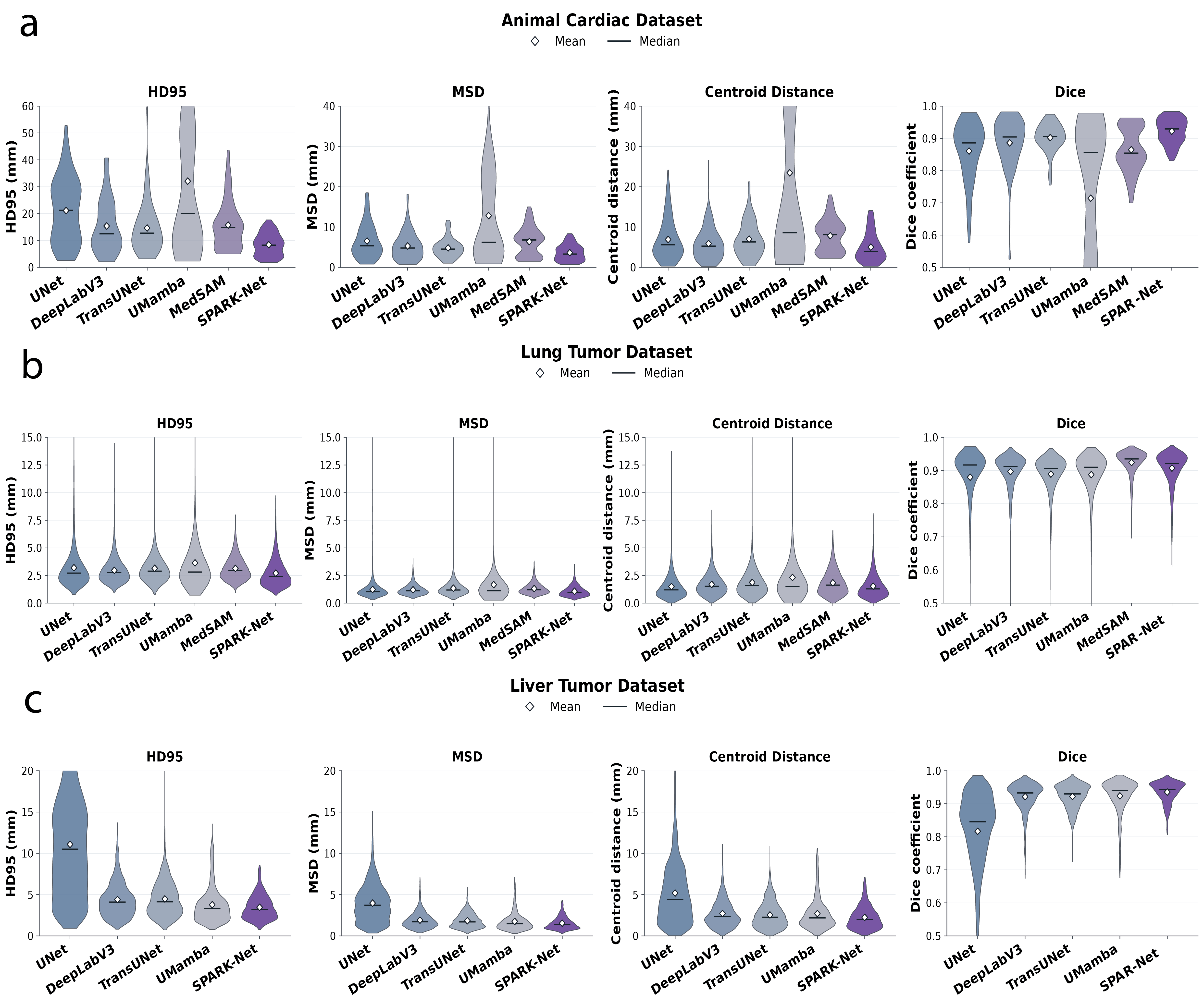}
    \caption{Violin plots illustrating the distributions of HD95, MSD, CD, and Dice across different methods for three datasets: (a) cardiac, (b) lung, and (c) liver. For improved visual clarity, the y-axis ranges are adjusted for each dataset to better illustrate the performance distributions.}
    \label{fig:jbhi_fig3}
\end{figure*}

\begin{figure*}[!t]
    \centering
    \includegraphics[width=0.85\textwidth]{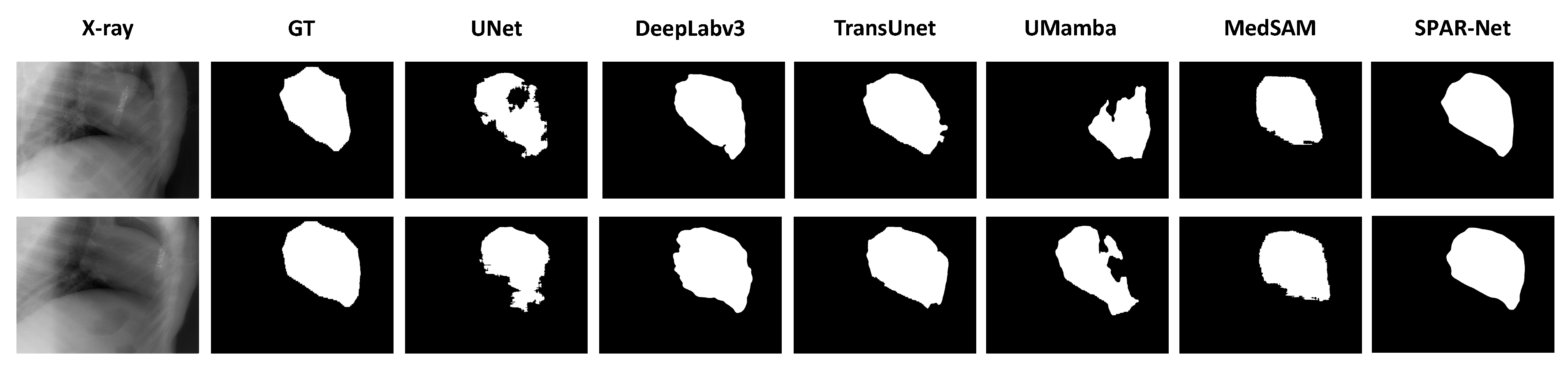}
    \caption{Visual comparison of SPAR-Net with existing state-of-the-art medical image segmentation methods using representative cases from the animal cardiac dataset.}
    \label{fig:jbhi_fig4}
\end{figure*}

\begin{figure*}[!t]
    \centering
    \includegraphics[width=0.85\textwidth]{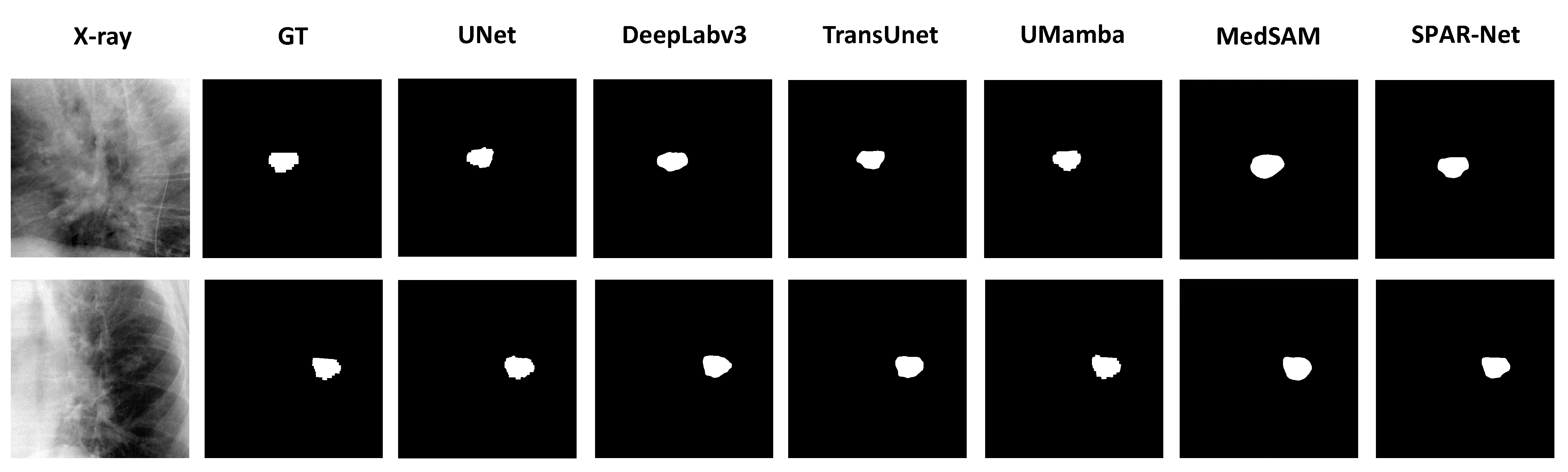}
    \caption{Visual comparison of SPAR-Net with existing state-of-the-art medical image segmentation methods using representative cases from the lung tumor dataset.}
    \label{fig:jbhi_fig5}
\end{figure*}

The lung tumor segmentation results are summarized in Table ~\ref{tab:lung_results}. SPAR-Net achieved the lowest overall HD95, MSD, and CD of 2.6 mm, 1.0 mm, and 1.5 mm, respectively. Compared with DeepLabV3, the best-performing baseline in terms of HD95, these corresponded to reductions of 13.3\%, 16.7\%, and 11.8\%. At the individual patient level, SPAR-Net consistently achieved the best boundary delineation and centroid localization accuracy across both patients. However, MedSAM achieved the highest overall Dice of 0.91, compared with 0.89 for SPAR-Net, despite having larger errors in all three distance-based metrics.
The violin plots in Fig.~\ref{fig:jbhi_fig3}b reveal further differences in performance variation. UNet, DeepLabV3, TransUNet, and UMamba exhibited extended upper tails in the distance-based metrics, indicating that a subset of projections resulted in substantially larger segmentation and localization errors. In contrast, SPAR-Net and MedSAM showed more concentrated distributions with fewer extreme errors. Notably, MedSAM used GT bounding-box prompts during inference, whereas SPAR-Net performed segmentation without spatial prompts, eliminating the need for prior knowledge of target location and enabling a more automated approach suitable for real-time image guidance.
The representative predictions in Fig. ~\ref{fig:jbhi_fig5} further illustrate these differences. Although all methods successfully identified the small target regions, variations in boundary shape, target size, and spatial position remained. MedSAM tended to produce slightly larger and more rounded masks, with some predicted pixels extending beyond the GT boundaries. In comparison, SPAR-Net generated compact masks that more closely followed the GT in shape and position. These observations are consistent with the quantitative results, where MedSAM achieved higher Dice but SPAR-Net obtained lower boundary and centroid errors. Overall, SPAR-Net maintained accurate target localization with fewer large-error predictions across both patients.

\subsection{Performance Comparison on the Liver Tumor Dataset}
\begin{figure*}[!t]
    \centering
    \includegraphics[width=0.85\textwidth]{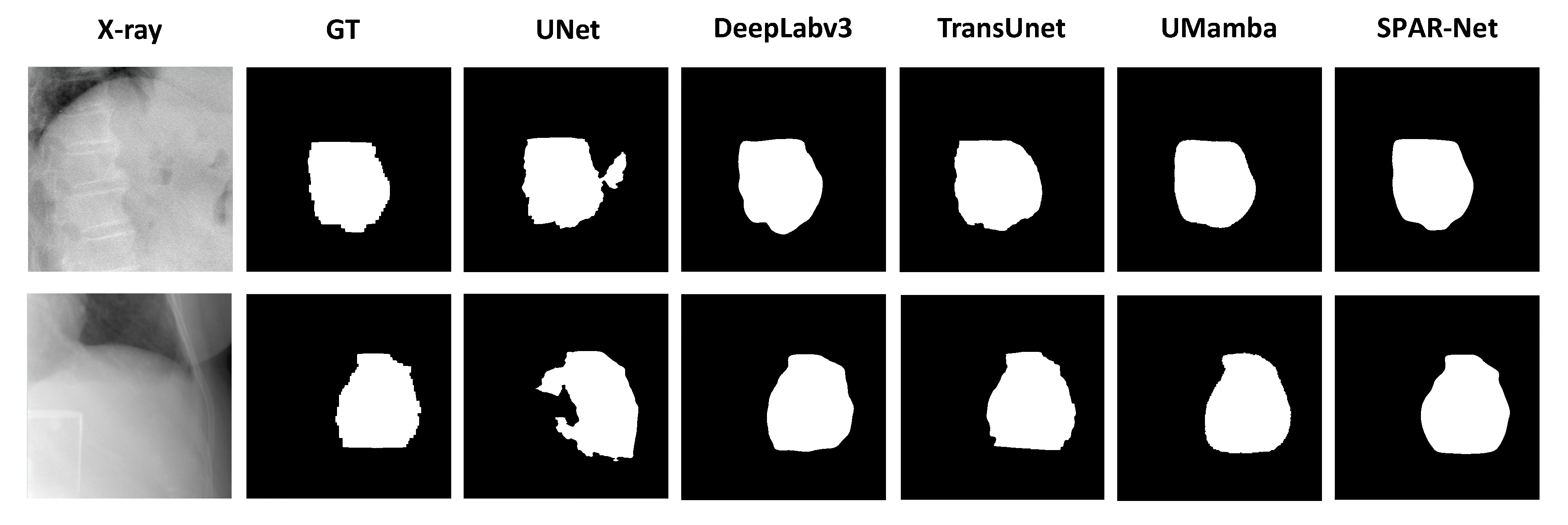}
    \caption{Visual comparison of SPAR-Net with existing state-of-the-art medical image segmentation methods using representative cases from the liver tumor dataset.}
    \label{fig:jbhi_fig6}
\end{figure*}

The liver tumor segmentation results are presented in Table ~\ref{tab:liver_results}. SPAR-Net achieved the best overall performance across all four metrics, with an HD95 of 3.2 mm, MSD of 1.4 mm, CD of 2.0 mm, and Dice Dice of 0.94. Compared with the strongest reported baseline values, SPAR-Net reduced HD95 by 5.9\%, MSD by 12.5\%, and CD by 13.0\%. It also achieved the lowest HD95 across all five patients, with the greatest patient-specific reduction of 23.1\% observed for PAT03.
The violin plots in Fig. 3c show substantial performance variation for UNet, with long upper tails in HD95, MSD, and CD. As liver tumors are not directly visible in X-ray images, their locations must be inferred from surrounding anatomical context. UNet struggled in this setting, producing large segmentation errors in some projections. In comparison, DeepLabV3, TransUNet, and UMamba showed more concentrated distributions, indicating more stable performance. SPAR-Net maintained a narrower distribution of localization errors, with fewer large deviations when estimating tumor positions from the surrounding anatomy.
The representative examples in Fig.~\ref{fig:jbhi_fig6} highlight differences in how the models estimated tumor boundaries and spatial extent. UNet frequently produced incomplete masks with irregular boundaries, including missing regions and inaccurate extensions beyond the GT. Although DeepLabV3, TransUNet, and UMamba generated more complete predictions, some differences in tumor size, shape, and position remained. In comparison, SPAR-Net more closely reproduced the GT contours, with fewer missing regions and better alignment of the predicted tumor locations. These differences were particularly evident in cases where the surrounding anatomical structures provided limited visual information for identifying the target.


\begin{table*}[!t]
\centering
\caption{Ablation study performed on the animal cardiac dataset. Results are reported as the mean across three animals. The last three columns report the percentage of X-ray projections with absolute centroid errors exceeding 5 mm in the X direction, Y direction, or either direction, using 5 mm as a clinically relevant localization threshold. The best scores are highlighted in bold.}
\label{tab:ablation}

\setlength{\tabcolsep}{3pt}
\renewcommand{\arraystretch}{1.3}

\resizebox{0.85\textwidth}{!}{%
\begin{tabular}{lccccccc}
\toprule

\multirow{2}{*}{Methods}
& \multicolumn{4}{c}{Segmentation Performance}
& \multicolumn{3}{c}{Centroid Distance Exceedance (\%)} \\

\cmidrule(lr){2-5}
\cmidrule(lr){6-8}

& HD95 (mm)
& MSD (mm)
& CD (mm)
& Dice
& X $>$ 5 mm
& Y $>$ 5 mm
& Either $>$ 5 mm \\

\midrule

Image branch
& $12.4\pm4.3$
& $5.2\pm1.9$
& $6.6\pm3.1$
& $0.90\pm0.05$
& 40.8
& 11.0
& 42.7 \\

Image branch + Domain Encoder
& $11.4\pm3.9$
& $4.2\pm1.9$
& $6.2\pm2.8$
& $0.91\pm0.04$
& 37.6
& 15.3
& 39.5 \\

SPAR-Net w/o random feature masking
& $9.2\pm3.5$
& $4.0\pm1.9$
& $5.6\pm2.5$
& $\mathbf{0.92\pm0.04}$
& 36.8
& 13.4
& 39.2 \\

SPAR-Net
& $\mathbf{8.7\pm3.3}$
& $\mathbf{3.8\pm1.8}$
& $\mathbf{5.3\pm2.3}$
& $\mathbf{0.92\pm0.04}$
& $\mathbf{36.3}$
& $\mathbf{11.6}$
& $\mathbf{38.7}$ \\

\bottomrule
\end{tabular}%
}

\end{table*}

\subsection{Ablation Study}
The ablation study was conducted on the cardiac dataset to evaluate the contributions of the domain encoder, physiological guidance, and random feature masking, as summarized in Table.~\ref{tab:ablation}. The image-only branch achieved an HD95 of 12.4 mm, MSD of 5.2 mm, CD of 6.6 mm, and Dice of 0.90. Incorporating the domain encoder improved all four metrics, reducing HD95 by 8.1\%, MSD by 19.2\%, and CD by 6.1\%, while increasing Dice to 0.91.
The full SPAR-Net achieved further improvements with the integration of physiological guidance. Compared with the image branch with domain adaptation, incorporating physiological information reduced HD95 by 23.7\%, MSD by 9.5\%, and CD by 14.5\%, while increasing Dice from 0.91 to 0.92. 
To further assess the effect of random feature masking within the MPSRU, we evaluated SPAR-Net with this mechanism disabled. Removing random masking increased HD95 from 8.7 to 9.2 mm, MSD from 3.8 to 4.0 mm, and CD from 5.3 to 5.6 mm, while Dice remained unchanged at 0.92. 

\section{Discussion}
In our study, we proposed SPAR-Net, a multimodal framework that integrates anatomical imaging with synchronized physiological signals for real-time motion estimation through target segmentation, using data available from existing imaging equipment. Across the cardiac, lung, and liver tasks, SPAR-Net demonstrated competitive performance across centroid distance, Dice coefficient and mean surface distance, with the greatest improvements observed in the 95th percentile Hausdorff distance (HD95), suggesting that physiological information can complement image-based representations for localizing moving anatomical targets. Mean HD95 values were 8.7, 2.7 and 3.2 mm for the three anatomical sites, representing reductions of 40\%, 10\% and 6\%, respectively, compared with the strongest competing methods (Tables~\ref{tab:cardiac_results}--\ref{tab:liver_results}). Image-level statistical tests further showed that SPAR-Net achieved statistically significant improvements across all evaluation metrics compared with the baseline methods. However, patient-level statistical testing was not performed due to the limited number of patients.

An important finding is that the improvements in boundary-based metrics were greater than those in Dice, particularly in the cardiac task. Motion estimation requires accurate localization of target displacement and deformation, rather than simply identifying the target region. Although Dice measures overall spatial overlap, it may be less sensitive to local boundary errors, particularly for relatively large anatomical targets where substantial overlap can be maintained despite noticeable contour deviations. In contrast, CD reflects positional errors, while HD95 and MSD provide complementary measures of boundary accuracy. Notably, reductions in HD95 indicate improved agreement in regions exhibiting larger boundary discrepancies, rather than improvements in average boundary accuracy alone. These geometric measures are particularly relevant to real-time image-guided radiotherapy, where anatomical motion, limited soft-tissue contrast and overlapping structures can introduce localized segmentation inaccuracies. Such errors may affect the estimated position and motion of anatomical structures relative to the treatment beam, potentially affecting treatment accuracy and increasing unintended irradiation of surrounding healthy tissues.

Among the baseline methods, MedSAM demonstrated competitive segmentation performance, particularly in terms of Dice. However, its reliance on GT bounding-box prompts presents a practical limitation for real-time radiotherapy applications, where target locations are not available during treatment sessions. This limitation becomes particularly relevant for targets that are poorly visible or invisible in X-ray projections, making reliable prompt generation challenging. DeepLabV3 and TransUNet also achieved competitive overall results, but exhibited greater variation across patients and individual projections, suggesting less consistent performance under different imaging conditions. Although TransUNet achieved segmentation accuracy relatively close to SPAR-Net, it required a longer inference time of 144 ms per image, compared with 116 ms for SPAR-Net (Tables~\ref{tab:cardiac_results}). This represents a reduction of 28 ms per image, or approximately 19.4\%. Computational efficiency is particularly important for real-time target tracking in radiotherapy, where imaging and treatment delivery operate on short timescales and delays may limit opportunities for motion correction. With an inference time of 116 ms per image, SPAR-Net demonstrates computational feasibility within the 500 ms latency requirement~\cite{ref50} for real-time image-guidance, although end-to-end system latency remains to be evaluated.

Differences between training and testing images introduce an additional challenge for real-time motion estimation in clinical settings. In a typical radiotherapy workflow, planning images, commonly acquired using CT, are obtained before treatment. However, imaging acquired during treatment may differ substantially in image contrast, and anatomical appearance, introducing a considerable domain shift. This challenge was particularly evident in the cardiac task, where the ablation study demonstrated the contribution of the SimSiam-pretrained ResNet-18 domain encoder, reducing the proportion of X-ray projections with centroid errors exceeding 5 mm in either the X or Y direction by 3.2 percentage points (Tables~\ref{tab:ablation}). Further integration of physiological guidance through the MPSRU and CRU reduced this proportion by another 0.8 percentage points, with random feature masking within the MPSRU contributing an additional 0.5-percentage-point reduction compared with the model without masking. Although these improvements appear modest, the combined 4.0-percentage-point reduction corresponds to approximately 20 fewer X-ray images exceeding the 5 mm threshold in a treatment sequence of 500 X-ray images. This reduction is clinically relevant, as large localization localization errors may result in the target extending beyond the planned treatment margin (typically 5 mm), potentially compromising target coverage or overlapping with nearby critical structures. Reducing the frequency of these errors could improve the reliability of real-time motion monitoring and support more accurate treatment delivery.

Beyond the cardiac task, anatomical variations between treatment sessions introduce additional challenges for lung and liver tumor localization, particularly when targets are poorly visible or invisible in X-ray projections. In these cases, the dilated convolutional decoder captures broader anatomical context, while the MPSRU integrates respiratory information to support target localization across different respiratory states. In addition, the CRU selectively weights feature channels based on physiological information, potentially emphasizing features associated with tumor motion and anatomical landmarks that remain informative across fractions. Together, these components may help establish correspondences between current X-ray images and anatomical features learned during training, even when tumor boundaries cannot be directly identified. 

Overall, SPAR-Net provides a scalable and accessible approach to real-time image guidance in radiotherapy and image-guided interventions. By integrating X-ray imaging with physiological signals available from existing clinical equipment, the framework may improve motion estimation without requiring additional costly imaging hardware such as MRI-Linacs, addressing an important clinical need, with 70\% of radiotherapy centers worldwide expressing interest in adopting real-time image guidance~\cite{ref51}. Furthermore, continuously acquired physiological signals may enable motion estimation between intermittent X-ray acquisitions, potentially reducing imaging frequency and associated radiation exposure. Previous work has suggested that X-ray imaging frequency could be reduced by up to 95\%~\cite{ref52}, although this was not evaluated in the present study and requires further investigation.

Several limitations remain. First, the number of patients was relatively small. Nevertheless, the dataset comprised over 14,000+ X-ray images across multiple anatomical sites, imaging sessions and conditions, with independently established GT. Further validation on larger, independent datasets is required to establish generalizability, potentially supported by the ongoing clinical trial (NCT05184790). Second, establishing GT anatomical positions in X-ray images is challenging for targets with poor soft-tissue contrast. In the lung and liver datasets, implanted gold fiducial markers provided GT tumor positions and were subsequently masked to prevent the model from relying on them. However, these markers are indirect surrogates that may not fully capture tumor motion and deformation, with reported localization uncertainties of 1–3 mm~\cite{ref53}.  Another limitation is that SPAR-Net performs segmentation in 2D X-ray projections, whereas respiratory and cardiac motion occurs in three dimensions. Consequently, motion along the X-ray beam direction cannot be directly resolved. Future work could integrate the estimated 2D anatomical positions with existing 2D-to-3D motion reconstruction methods~\cite{ref54}, potentially incorporating multiple imaging angles or patient-specific anatomical models to enable three-dimensional motion estimation.

Our study provides a general framework for integrating X-ray imaging with physiological signals, demonstrated using ECG and respiratory surface signals. In the future, this framework could be extended to incorporate additional modalities, such as electromyography (EMG) and electroencephalography (EEG). EMG could capture respiratory muscle activation preceding mechanical motion, while EEG may provide information about neural activity associated with voluntary movement. Integrating these complementary signals could further improve real-time motion estimation and enable prediction of future anatomical motion.

Future work will also focus on prospective multicenter clinical evaluation in our multi-institutional study (NCT05184790) to assess generalizability and support real-time clinical deployment. Uncertainty-guided approaches, including agreement between multiple models, could help identify unreliable predictions and support safer clinical decision-making during treatment.

\section{Conclusion}
In this study, we proposed SPAR-Net, a multimodal framework integrating X-ray imaging and physiological signals for anatomical motion estimation in radiotherapy and interventional imaging. Evaluated across cardiac, lung and liver sites, the framework demonstrates the potential of multimodal learning to enable accurate target motion estimation and boundary accuracy under challenging X-ray imaging conditions. This study provides a step towards scalable and accessible image guidance by leveraging physiological information to support real-time motion estimation.


\section*{References}

\bibliographystyle{IEEEtran}
\bibliography{references}

\end{document}